\documentclass[twocolumn,dvipsnames]{aastex631}
\usepackage{amsmath}
\usepackage{wrapfig}
\usepackage{}
\usepackage{soul}
\usepackage{placeins}

\usepackage{comment}

\begin{document}
\title{General form for Pseudo-Newtonian Potentials, imitating Schwarzschild geodesics}

\author[0009-0003-1403-5183]{Itamar Ben Arosh Arad}
\email{itamar.benarosh@mail.huji.ac.il}
\author[0000-0002-1084-3656]{Re'em Sari}
\affiliation{Racah Institute of Physics,Hebrew University of Jerusalem,9190501,Israel}


\begin{abstract}
We propose a new, general form for a pseudo-Newtonian gravitational potential (PNP), expressed as a series of Paczyński-Wiita-like functions with the addition of increasing negative powers of $r$ with arbitrary coefficients. We present a procedure for determining these coefficients to construct a custom PNP  that replicates key features of Schwarzschild geodesics for a test particle near a black hole.  As an example, we construct potentials 
set to reproduce (I) the presence of an innermost stable circular orbit at the $r=6$ (geometric units), with the correct infall velocity for small deviations (on the geodesic universal infall), (II) the periapsis advance at large distances, and (III) the presence of a marginally bound circular orbit with specific angular momentum $L=4$, and the periapsis advance of parabolic orbits close to it.
We compare the performance of our examples against the Paczyński-Wiita potential and other existing potentials. Finally, we discuss  the limitations and advantages of our formulation.

\end{abstract}

\section{Introduction}
In astrophysical high-energy phenomena,  such as tidal disruption events or accretion disks around black holes, general relativity plays an important role. However, in some cases, using relativistic dynamics in curved space-time greatly complicates the analysis - numerical or analytical alike.  Therefore, some research employs a pseudo-Newtonian approach: Newtonian dynamics with a central potential that is constructed to mimic key relativistic behaviors.

 \cite{1980A&A....88...23P} pioneered this approach, proposing a new  pseudo-Newtonian potential (PNP) known as the Paczyński-Wiita (PW) potential:
\begin{equation}
    \Phi_{pw}(r)=-\frac{GM}{r-r_{s}}\label{PWp},
\end{equation}
where $M$ is the black hole mass, $G$ is Newton's gravitational constant and $r_s\equiv{2r_{*}\equiv} \frac{2GM}{c^{2}}$ is the Schwarzschild radius.
This potential reproduces several essential features of orbits in Schwarzschild space-time. Specifically, it reproduces the innermost stable circular orbit (ISCO) radius at $3r_{s}$ and the marginally bound circular orbit (MBCO) radius at $2r_{s}$. It also yields the exact specific angular momentum value for any parabolic trajectories as shown in the work of \cite{abramowiczPaczynskiWiitaPotentialStepbystep2009a}.
This idea inspired several alternative PNPs. \cite{nowakDiskoseismologyProbingAccretion1991a}  introduced one that reproduces the Keplerian frequency of circular orbits outside the ISCO.
\cite{weggPSEUDONEWTONIANPOTENTIALSNEARLY2012a} proposed another potential (shown in geometric units $G=c=1$):\begin{equation}
    \Phi_{Wegg}=-\frac{\alpha M}{r}-\frac{(1-\alpha)M}{r-R_{x}}-\frac{R_{y}M}{r^{2}} \label{WeggPot}
\end{equation}
 where
 \footnotesize 
\begin{equation}\label{eq:Wcoef}
 \alpha=-\frac{4}{3}\left(2+\sqrt{6}\right),R_{x}=\left(4\sqrt{6}-9\right)M,\\
 R_{y}=-\frac{4}{3}\left(2\sqrt{6}-3\right)M
 \end{equation}
 \normalsize
This potential reproduces parabolic orbital precession per orbit with $<1$\% relative error, for any value of angular momentum. The coefficients in Equation \eqref{eq:Wcoef} were obtained by imposing constraints on the general form shown at Equation \eqref{WeggPot} to match key relativistic precession features.
\cite{tejedaAccurateNewtonianDescription2013a}  introduced a PNP that depends on both position and velocity, achieving high accuracy in reproducing relativistic geodesics. This approach of velocity-dependent PNPs has been addressed in several other works such as \cite{ghoshSphericalAccretionSchwarzschild2025} and \cite{witzanyPseudoNewtonianEquationsEvolution2017a}.
However, velocity dependence complicates the equations of motion. Therefore, we restrict this study to purely position‐dependent PNPs. Several other PNPs regarding Schwarzschild space-time and other space-time geometries such as Kerr, Schwarzschild–de Sitter, and Kerr–Newman, appear in 
\cite{mukhopadhyayDescriptionPseudoNewtonianPotential2002a}, \cite{mukhopadhyayPseudoNewtonianPotentialsDescribe2003a}, \cite{ivanovPseudoNewtonianPotentialCharged2005a}, \cite{stuchlikPseudoNewtonianGeneralRelativistic2009a},\cite{dihingiaLimitationsPseudoNewtonianApproach2018},\cite{shakuraLogarithmicPotentialGravitational2018}.

PNPs have seen extensive use in numerical simulations. \cite{kluzniakSwallowingQuarkStar2002a} modeled a quark-star–black-hole merger with a custom PNP.
Other previous studies such as \cite{crispinoPseudoNewtonianPotentialsRadiation2011a} , \cite{taniguchiInnermostStableCircular1996a}, \cite{hawleyHighResolutionSimulationsPlunging2002a} also use this method to study radiation by a source swirling around a stellar object, coalescing neutron star-black hole binaries and accretion disks. 

More recently \cite{steinbergStreamDiskShocks2024} applied the PW potential to simulate tidal disruption events and predict their light curves. Similarly, \cite{papavasileiouSimplifiedApproachReproducing2025},\cite{qiaoEarlyEvolutionSuperEddington2025},\cite{paulGravityEmergentPhenomenon2025},\cite{cemeljicPseudoNewtonianSimulationThin2025},\cite{wenPseudoNewtonianStationaryCircumbinary2025} also employ pseudo-Newtonian methods to study X-ray binaries, TDEs, accretion flow, accretion disks and circumbinary disks.
Each potential captures only certain relativistic effects and may deviate significantly in others. Depending on the phenomenon, some effects are crucial to capture, motivating the development of adaptable PNPs tailored to specific requirements.

In this paper, we propose a general pseudo-Newtonian potential, designed to reproduce various key features of test-particle orbits in Schwarzschild spacetime. The potential includes arbitrary coefficients that can be constrained to match selected general relativistic properties. We demonstrate this approach and compare its performance with that of previously proposed PNPs.

\section{The General PNP Form}

We present a new PNP composed of a series of PW-like terms and an independent series of terms proportional to $r^{-n}$:

\begin{equation}
\Phi_{pn}(r)=-\sum_{n=0}^{N_1-1}\alpha_{n}\frac{GMr^{n}}{(r-r_n)^{n+1}}-\sum_{n=0}^{N_2-1}\beta_{n}\frac{GM r_*^n}{r^{n+1}}\label{GenPot}
\end{equation}

The parameters $r_n$, $\alpha_n$, and $\beta_n$ constitute $2N_1+N_2$ arbitrary coefficients ($\alpha_n,\beta_n$ are dimensionless and $r_n$ has dimensions of length), and are selected to reproduce specific features of Schwarzschild geodesics. This family of potentials is constructed to encompass previously proposed PNPs (e.g., \citet{weggPSEUDONEWTONIANPOTENTIALSNEARLY2012a}, \citet{nowakDiskoseismologyProbingAccretion1991a}, and \citet{1980A&A....88...23P}) by retaining the original PW idea of shifting the divergence radius of the potential and by introducing additional negative powers of $r$, which provide improved control over the behavior of the potential at radii larger than the divergence radius.
This paper surveys a subgroup of these potentials where the dimensional coefficients are set to $r_n=const=r_s$, $r_s$ being the Schwarzschild radius.
This more limited version has  $N_1+N_2$ dimensionless coefficients, with the advantage that the potential is linear in all these free coefficients.
From this point onward, we adopt geometric units (${c=G=1}$) and set $M=1$ for simplicity, so that $r_s=2$. 
In these units Equation \eqref{GenPot}, becomes:
\begin{equation}
\Phi_{pn}(r)=-\sum_{n=0}^{N_1-1}\alpha_{n}\frac{r^{n}}{(r-2)^{n+1}}-\sum_{n=0}^{N_2-1}\beta_{n}r^{-n-1}\label{OurPot}
\end{equation}
Based on this form we proceed in three steps. First, we identify the features of Schwarzschild geodesics that are relevant to the system under consideration, such as the ISCO radius. Next, we use classical mechanics and the general form of the potential given by Equation \ref{OurPot}  to derive expressions for these features in terms of the coefficients, thereby yielding a system of linear equations. Finally, we solve this system of linear equations and determine the coefficients. This procedure requires that the number of coefficients, 
$N_1+N_2$, is  equal to the number of equations which reflect the desired potential properties.
For a given $N$, the partition between $N_1$ and $N_2$ could be arbitrary. 

An example for a fundamental feature of Schwarzschild space-time is that, at large radii $(r\gg 2 )$, the potential reduces to the Newtonian form $-1/r$,  consistent with the far-field limit of the Schwarzschild solution.
Matching the leading-order expansion of Equation \ref{OurPot} to the Newtonian potential yields the condition:
\begin{equation}
\sum_{n=0}^{N}\alpha_{n}+\beta_{0}=1\label{1stCond}
\end{equation}

We first define the effective potential for test particles in our pseudo-Newtonian framework and in the Schwarzschild spacetime.
For test particles with specific angular momentum $L$, and specific energy $E$,
the effective potential in the PNP framework is:
\begin{equation}
V_{eff}=\Phi_{pn}(r)+\frac{L^{2}}{2r^{2}}\label{EffPot}   
\end{equation}
For comparison, the GR effective potential in the Schwarzschild space-time  is defined as:
\begin{equation}
V_{eff}^{GR}=-\frac{2}{r}+\frac{L^{2}}{r^2}-\frac{2L^{2}}{r^3}\label{EffPotGR}   
\end{equation}
From these effective potentials, the radial equations of motion (EOMs) can be derived. In the GR case:
\begin{equation}
    \dot{r}=\frac{(1-\frac{2}{r})}{E}\sqrt{E^2-1-V_{eff}^{GR}(r,L)}
    \label{EOMgr}
\end{equation}
and in the Newtonian case:
    \begin{equation}
    \dot{r}=\sqrt{2E-2V_{eff}(r,L)}
    \label{EOMpn}
\end{equation}
Where $V_{eff}$ and $V_{eff}^{GR}$ are taken from Equations \eqref{EffPot} and  \eqref{EffPotGR}  respectively.

In the following sections, we give a few examples of what could be desired properties of the PNP, and calculate the corresponding requirement on the free coefficients $\alpha_n$ and $\beta_n$. We give a few examples of behavior around the Innermost Stable Circular Orbit (ISCO) in \S\ref{sec:ISCO} and a few examples of precession at large distances and close to the marginally bound orbit in \S\ref{sec:prec}.

\section{The ISCO Geodesic\label{sec:ISCO}}

In the Schwarzschild spacetime, stable circular orbits  are not possible for radii smaller than $r=3r_{s}=6$. The orbit at $r=6$  is thus known as the innermost stable circular orbit (ISCO). Equivalently, the ISCO corresponds to a stationary inflection point of the effective potential at $r=6$. 
This condition can be translated to the following equations: \begin{eqnarray} V'_{eff}(r=6,L=L_{ISCO})=0,\nonumber \\ V''_{eff}(r=6,L=L_{ISCO})=0. \label{2ndCondabstract}
\end{eqnarray}
Here, and throughout the paper, the prime symbol ($'$) denotes differentiation with respect to the radial coordinate $r$.
The first  equation yields an expression for the angular momentum of a circular orbit at $r=6$,  substituting this expression into the second equation gives the condition:
\begin{eqnarray}
 \sum_{n=0}^{N_{1}-1}\frac{1}{4}\alpha_{n}\left(\frac{3}{2}\right)^{n+1}\left(n^{2}+5n\right)
 \nonumber\\+\sum_{n=0}^{N_{2}-1}\left(n^{2}-1\right)\beta_{n}6^{-n}=0\;\label{2ndCond}   
\end{eqnarray}
This condition  ensures that  the ISCO appears at the correct radius, but it alone may not reproduce all features of the geodesic, such as the orbital frequency of the ISCO, the specific orbital energy or angular momentum  or the radial velocity during the geodesic universal infall (GUI). 
Therefore, additional constraints can be imposed to capture those remaining aspects.

One feature of interest is the infall rate of the GUI near the ISCO radius. The ISCO is marginally stable: any infinitesimal change to the particle's energy causes a plunge into the black hole along a trajectory known as the GUI (\cite{romExtremeMassRatioBinary2022a}). Modeling this plunge may be important for understanding the inner-edge dynamics of accretion disks and their associated high-energy emission.  Close to the ISCO radius, the particle descends in quasi-circular orbits with a small radial velocity. Capturing the full GUI trajectory is beyond the scope here, but the leading-order relativistic radial velocity (Equation \eqref{EOMgr}) near $r=6$ on the GUI can be reproduced.
From Equation \eqref{2ndCondabstract}, the first and second order terms in the Taylor expansion of the effective potential vanish.  Consequently, the leading-order radial velocity equation reduces to:
\begin{equation} \dot{r}\approx{\sqrt{\frac{1}{2}}}\sqrt{-\frac{1}{3!}V_{eff}'''^{GR}(r=6,L=\sqrt{12})(r-6)^{3}}
    \label{EOMgrapprox}.
\end{equation}
A similar expansion may be applied to the Newtonian radial velocity (Equation \eqref{EOMpn}) for a general potential.
The specific angular momentum and energy are given by $L_{ISCO}=6^{3}\frac{d}{dr}\Phi(r)|_{r=6}$ and $E=\frac{1}{2}V_{eff}(6;L_{ISCO})$, respectively.  Substituting these into Equation \ref{EOMpn} yields:
\begin{equation}
\dot{r}\approx\sqrt{-\frac{1}{3!}V'''_{eff}(r=6,L=L_{ISCO})(r-6)^{3}}.\label{EOMpnapprox}
\end{equation}
Equating the two expansions implies that matching the leading-order infall behavior requires:
\begin{equation}
V'''_{eff}(r=6,L=L_{ISCO})=\frac{1}{2}V_{eff}'''^{GR}(r=6,L=\sqrt{12})
\end{equation}
Substituting the general potential form (Equation \eqref{OurPot}) and evaluating the resulting derivatives then yields the next condition:
\begin{eqnarray}
 \sum_{n=0}^{N_{1}-1}\frac{\alpha_{n}}{8}\left(\frac{3}{2}\right)^{n+1}(18+89n+24n^{2}+n^{3})
\nonumber\\+\sum_{n=0}^{N_{2}-1}\left(n-1\right)\left(n+1\right)\left(n+6\right)\frac{\beta_{n}}{6^{n}}=1 \label{4thCond}  
\end{eqnarray}
Condition \eqref{2ndCond}  is necessary for the existence of an ISCO, but not sufficient.

For an ISCO to exist, the effective potential must have no additional extrema at the same angular momentum aside from the stationary inflection point at $r=6$.
Ensuring that the ISCO is the only extremum amounts to requiring that the effective potential is strictly increasing for all $r\neq6$. As a result, determining the coefficients becomes a constrained optimization problem, rather than a straightforward solution of linear equations.
The divergence of the effective potential close to the Schwarzschild radius demonstrates how certain parameter choices can break monotonicity. By definition the effective potential diverges as $r\rightarrow 2$, but whether it tends to $+\infty$ or $-\infty$  depends on the sign of $\alpha_{N1}$. As one can see:
\begin{equation}
V_{eff}(r\rightarrow 2;L)\approx-\alpha_{N_{1}}\frac{2^{N_{1}}}{(r-2)^{N_{1}+2}}|_{r\rightarrow 2}+Const\label{prob_number_2}
\end{equation}
As seen in Equation \eqref{prob_number_2}, if $\alpha_{N1}<0$, the effective potential diverges to $+\infty$, preventing any plunge trajectory and possibly allowing an additional stable circular orbit inside the ISCO. Thus, even if the coefficients satisfy conditions \eqref{2ndCond} and \eqref{4thCond}), the ISCO may still fail to exist in the usual sense.


\section{precession rates\label{sec:prec}}
Orbital precession is another essential feature of Schwarzschild geodesics, characterized by a gradual shift in its orbital periapsis with each revolution. The precession angle per orbit is given by:
\begin{equation}
    \Delta\phi=2\int _{r_{\boldsymbol{-}}}^{r^{+}}\frac{L}{r^{2}\sqrt{E^2-(1-\frac{2}{r})(1+\frac{L^{2}}{r^{2}})}}
dr-2\pi\label{PrecCalcGR}
\end{equation}
where  $r_{\pm}$ are the periapsis and apoapsis, determined by the equations:
\begin{equation}
E^{2}-(1-\frac{2}{r_{\boldsymbol{\pm}}})(1+\frac{L^{2}}{r_{\boldsymbol{\pm}}^{2}})=0 
\label{PriGR}
\end{equation}
 In a Newtonian framework with a general potential, one obtains the analogous expression
\begin{equation}
\Delta\phi=2\int _{r_{\boldsymbol{-}}}^{r^{+}}\frac{L}{r^{2}\sqrt{2E-2V_{eff}(r,L)}}
dr-2\pi\label{PrecCalcPNP}
\end{equation}
and parallel to \eqref{PriGR}:
\begin{equation}
E-V_{eff}(r_{\boldsymbol{\pm}})=0\label{PriPNP}
\end{equation}
Where $V_{eff}$ is defined as in Equation \eqref{EffPot}.
The integral in Equation \eqref{PrecCalcGR}  cannot be evaluated in closed form for arbitrary $(E,L)$ and must instead be computed numerically.  Two asymptotic regimes, however, admit analytic approximation and can serve to constrain the PNP coefficients.
The first case concerns the leading-order behavior of precession angle for orbits with large periapsis distances. Following the derivation of \cite{weinbergGravitationCosmologyPrinciples1976}, the precession per orbit for an eccentric orbit in a spherically symmetric metric, at the limit where $r_{-}\gg2$ is given by the expansion:
\begin{equation}
\Delta\phi= 0 + 3\pi(\frac{1}{r_{+}}+\frac{1}{r_{-}})+\mathcal{O}(\frac{1}{r_{-}^{2}})\label{GRfarperc}
\end{equation}
Expanding the expression for $L^2$ with respect to $r_\pm$ shows that $L^{-2}\approx\frac{1}{2}(r_-^{-1}+r_+^{-1})$, substituting this relation into \eqref{GRfarperc} yields:
\begin{equation}
\Delta\phi\approx \frac{6\pi}{L^2}\label{GRfarpercL}
\end{equation}
This relation also holds for parabolic orbits, in which the apoapsis lies at infinity ($r_{+}=\infty$). The $0th$ order vanishes, since in the classical Newtonian case there is no precession.
Using a similar derivation in the Newtonian case (using Equations \eqref{GenPot},\eqref{EffPot} and \eqref{PrecCalcPNP}), leads to the expression:
\begin{equation}
\Delta\phi= 0 + (\sum_{n=0}^{N_{1}-1}\alpha_{n}2\left(n+1\right)+\beta_{1})\pi(\frac{1}{r_{+}}+\frac{1}{r_{-}})+\mathcal{O}(\frac{1}{r_{-}^{2}}) \label{Pnfarprec}
\end{equation}
The same approximation of the relation between $L^2$ and $r_\pm$ applies in the Newtonian case as well.
This leads to the following condition:
\begin{equation}
2\sum_{n=0}^{N_{1}-1}\alpha_{n}\left(n+1\right)+\beta_{1}=3\label{5thCond}  
\end{equation}

The second asymptotic regime arises as $L\rightarrow 4^+$. In this limit, the precession per orbit diverges logarithmically. To simplify the analysis, parabolic orbits ($E=1$) are considered. This divergence may be understood via the geometry of parabolic orbits. In such an orbit, the precession corresponds to a net rotation of the parabola’s axis of symmetry. As $L\rightarrow 4^+$, the particle executes an increasing number of loops at radii near $r=4$, accumulating an extra $2\pi$ of azimuthal advance with each loop. Consequently, the precession grows without bound, following a logarithmic divergence as $\delta L= (L-4)\rightarrow 0$.

Setting $E=1$ in Equation \eqref{PrecCalcGR} allows for an examination of the leading order behavior of $\Delta\phi$ as $L$ approaches $4$. The leading order behavior found, as also shown in \cite{weggPSEUDONEWTONIANPOTENTIALSNEARLY2012a} is:
\begin{equation}
    \lim_{L\rightarrow4}\Delta\phi=-\sqrt{2}\ln(L-4)\label{PrecGrL4}
\end{equation}
The specific energy of  a parabolic orbit differs between Newtonian mechanics and general relativity: in the Newtonian case, it is zero, while in GR it is one due to the inclusion of rest mass energy. In geometric units, the relation between the relativistic and Newtonian specific energies is $E_{gr}^2=E_n+1$.

To begin, the pseudo–Newtonian potential must reproduce the MBCO radius at $r=4$, corresponding to  $L=4$. The  MBCO is found by requiring both the first derivative and the value of the effective potential to vanish at that MBCO radius. This can be shown as
\begin{eqnarray}
    V_{eff}'(r=4,L=L_{MBCO})=0\nonumber \\V_{eff}(r=4,L=L_{MBCO})=0\label{6th7thabstract}
\end{eqnarray}
Substituting the proposed PNP into Equation (\ref{6th7thabstract}) yield
s: 
\small
\begin{equation}
    2\sum_{n=0}^{N_{1}-1}{2^{n}}n\alpha_{n}+\sum_{n=0}^{N_{2}-1}\left(n-1\right)4^{-n}\beta_{n}=0\label{6thCond}
\end{equation}
\normalsize

Additionally, the pseudo–Newtonian angular momentum can be constrained to match the general relativistic value $L=4$.
Substituting $L_{MBCO}=4$ into one of the equations in  Equation \eqref{6th7thabstract} yields the additional constraint:
\begin{equation}
    \sum_{n=0}^{N_{1}-1}2^{n}(4+2n)\alpha_{n}+\sum_{n=0}^{N_{2}-1}\left(n+1\right)4^{-n}\beta_{n}=4\label{7thCond}
\end{equation}

Once the existence of the MBCO is established, the next step is to ensure that the PNP reproduces the logarithmic divergence of the precession as $L$ approaches $4$ from above. As shown in Equation \eqref{PrecCalcGR} and Equation \eqref{PrecCalcPNP}, The precession angle for a parabolic orbit is calculated by:

\begin{equation}
 \Delta\phi=2\int _{r_{\boldsymbol{-}}}^{\infty}\frac{L}{r^{2}\sqrt{-2V_{eff}}}dr-2\pi\label{ParaboliPrecCalc}
\end{equation}
Also from Equation \eqref{PriGR} and Equation \eqref{PriPNP}, another relation is obtained :
\begin{equation}
-V_{eff}(r_{-})=0\label{ParaboliPriCalc}
\end{equation}
The effective potential in the  Newtonian mechanics case and in the GR case  are both approaching $0$ from below as $r$ approaches infinity at the same rate (This is equivalent to saying that the physical behavior of the system becomes classical Newtonian gravity for large radii, also demanded in Equation \eqref{1stCond}). Therefore, the dominant contribution to the integral in Equation \eqref{ParaboliPrecCalc}   arises from the region near $r_{-}$ , where the integrand diverges. Equation \eqref{ParaboliPrecCalc} can thus be approximated as:
\begin{equation*}
 \Delta\phi\approx2\frac{L}{r_{-}^{2}}\int _{r_{\boldsymbol{-}}}^{r_{\boldsymbol{-}}+a}\frac{1}{\sqrt{-2V_{eff}}}dr+C,
\end{equation*}
where $r_{-}+a$ is an arbitrary cutoff point. 
The MBCO corresponds to a local maximum of the effective potential. A small change in angular momentum, such as $L=4+\delta L$ where $\delta L\ll 1$, produces a slight elevation of that maximum proportional to $\delta L$, and shifts the periapsis to the right by an amount proportional to  $\sqrt{\delta L}$.  This variation also causes only a negligible change in the curvature of the maximum and its radial coordinate. Since $a$ is arbitrary, it may be chosen sufficiently small, allowing the effective potential near the periapsis to be approximated by a parabola.
Using this approximation, the integral evaluates to:
\begin{equation}
 \Delta\phi\approx -4\sqrt{-2V_{eff}''(r=4,L=L_{MBCO})}\ln(L-4)+C
\end{equation}
Reproducing the divergence behavior of Equation \eqref{PrecGrL4}, requires: $4\sqrt{-V_{eff}''(r=4,L=L_{MBCO}4)}=1$.
Substituting the PNP effective potential into this expression yields the following condition:"
\begin{equation}
    2\sum_{n=0}^{N_{1}-1}2^{n}\left(n^{2}+4n+2\right)\alpha_{n}+\sum_{n=0}^{N_{2}-1}4^{-n}\left(n^{2}-1\right)\beta_{n}=2
    \label{8thCond}
\end{equation}
The condition above is independent of condition \eqref{7thCond}, as this was achieved
by using a general expression for $L_{MBCO}$ at $r_{-}=4$. 
This allows the condition to be used without the need to also impose \eqref{7thCond}.

Improved accuracy in orbital precession is obtained by matching the third derivative of the effective potential at $L=L_{MBCO},r=4$ to the GR counterpart. This additional constraint improves the local approximation of the effective potential near $r=4$. The resulting condition is:
\begin{eqnarray}
    \sum_{n=0}^{N_{1}-1}2^{n-2}(192+352n+120n^{2}+8n^{3})\alpha_{n}+\nonumber\\ \frac{1}{2}\sum_{n=0}^{N_{2}-1}4^{-n}\left(n-1\right)\left(n+1\right)\left(6+n\right)\beta_{n}=9\label{9thCond}
\end{eqnarray}
This condition above is also independent of condition \eqref{7thCond}.

The conditions above are just a few examples of the constraints that can be imposed on the general potential form (Equation \eqref{GenPot}). Additional examples include matching higher derivatives of the effective potential at key points, such as at the MBCO radius, equating the specific angular momentum or orbital energy of selected orbits, reproducing the light‑ring radius, matching orbital frequencies, and so on.
In principle, any number of constraints can be applied to this potential form. However, an arbitrary set of conditions may not yield a solvable system of equations, and it may be impossible to find a set of coefficients that satisfies all the conditions.

\begin{figure}[!]
    \centering
    \includegraphics[width=1\linewidth]{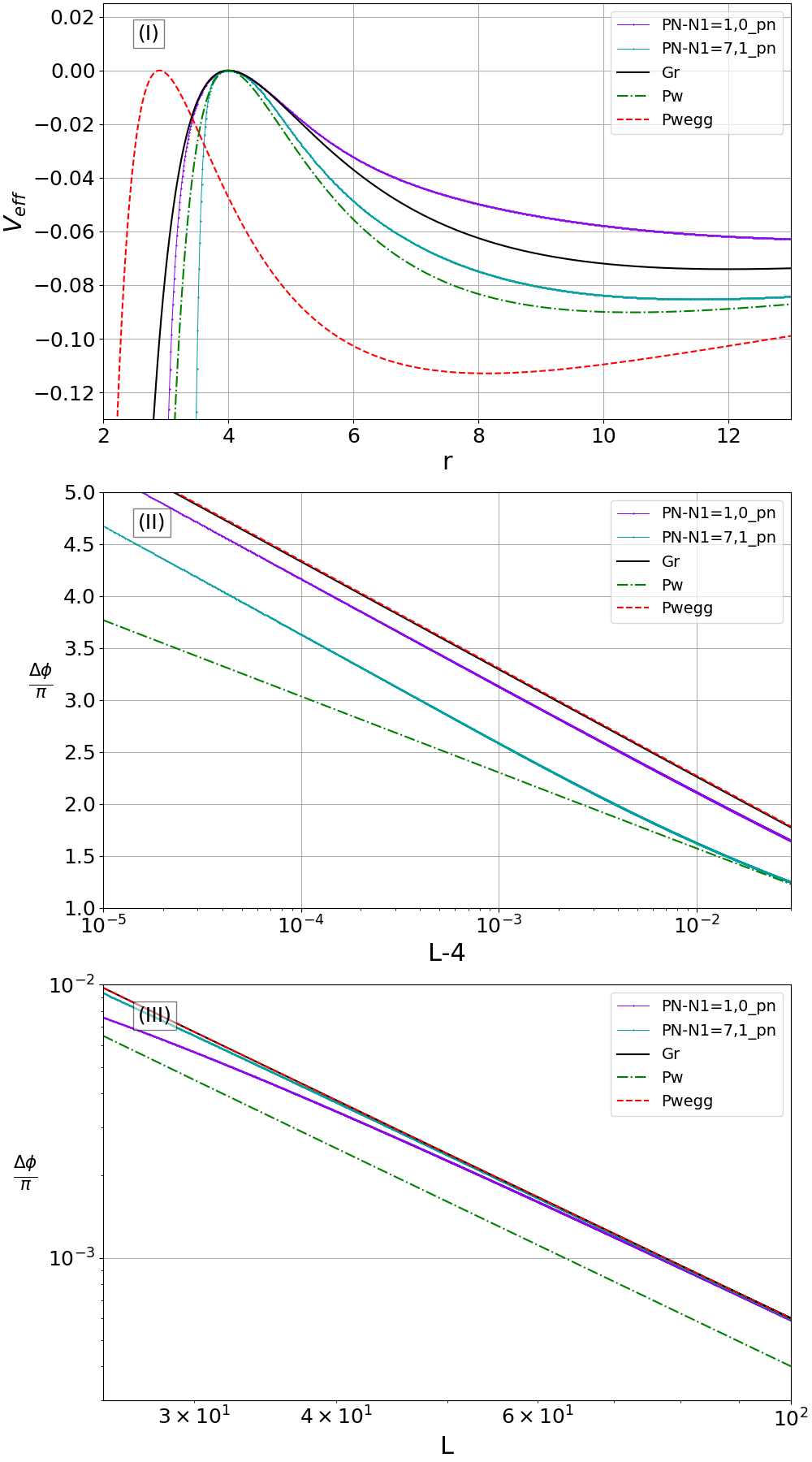}
    \caption{\footnotesize
    Comparison of two pseudo‑Newtonian potentials with $N_1=7$ (purple line) and $N_1=1$ (turquoise line), constructed using the constraints presented in sections \S\ref{sec:ISCO} and \S\ref{sec:prec}, alongside the PW potential (Equation \eqref{PWp}, green dotted line), Wegg's potential (Equation \eqref{WeggPot}, red dashed line)  and the exact GR results (solid black line). 
    Panel (I): Effective potential versus $r$ for $L=4$, corresponding to the angular momentum of the marginally bound circular orbit.
    Panel (II): Precession per orbit (in units of $\pi$) for a parabolic trajectory, plotted against  $L-4$ in logarithmic scale, illustrating the logarithmic divergence as $L \rightarrow 4^+$.
    Panel (III): Precession per orbit  (in units of $\pi$) for a parabolic trajectory, versus $L$ in the far field regime ($L\gg4$, log‑log scale), demonstrating the $\frac{1}{L^2}$ scaling of the leading correction to the precession.}
    \label{fig:Fig1}
\end{figure}
\section{Examples and Comparison}\label{sec:exm}

The general form of the potential in Equation \eqref{OurPot} can yield many PNPs that reproduce Schwarzschild geodesics with varying accuracy.  Here, as an example, we construct two potentials by imposing the eight conditions presented in the previous sections. In one we set $N_1=1,N_2=7$ and in the second we set $N_1=7,N_2=1$. The conditions we impose are: reproducing the far‑field convergence to classical Newtonian gravity \eqref{1stCond},  reproducing the ISCO radius at $r=6$ \eqref{2ndCond}, matching the infall rate of the GUI as $r\rightarrow 6$ \eqref{4thCond},  reproducing the leading‑order precession per orbit in the far‑field limit  \eqref{5thCond}, setting the MBCO radius to $r=4$ \eqref{6thCond}, fixing the angular momentum of that orbit to $L=4$ \eqref{7thCond}, and capturing the logarithmic divergence of the precession per orbit as $L \rightarrow 4^+$ (\ref{8thCond} and \ref{9thCond}). Solving this system of equations determines the free coefficients and the two solutions are listed in Table \eqref{tab:coefficients_vertical}.

Figure [\ref{fig:Fig1}] presents a comparison of these PNPs, based on the $L=4$ effective potential and the precession per orbit. In this figure, the precession per orbit was calculated numerically by evaluating the full integrals in Equation \eqref{PrecCalcGR} and Equation \eqref{PrecCalcPNP}.
Both PNPs successfully reproduce the correct MBCO radius, in contrast to Wegg’s potential. In panel [\ref{fig:Fig1}(I)], the effective potential has a maximum point at $(4,0)$ for both PNPs and the PW potential, while Wegg’s potential reaches its maximum at $(2\sqrt{6}-2,0)$, indicating a different MBCO radius. 
 Furthermore, the PNPs capture the first‑order orbital precession in both limiting regimes, as shown in panels [\ref{fig:Fig1}(II)] and [\ref{fig:Fig1}(III)].
For angular momentum near $L=4$, recalling Equation \eqref{PrecGrL4}, a semi‑log plot displays the logarithmic divergence as a linear plot whose slope gives the divergence amplitude. Panel [\ref{fig:Fig1}(II)] shows that both PNPs become parallel to the GR line as $L\rightarrow 4^+$, indicating they have the correct amplitude, whereas the PW potential does not. Computing the slope in the region $L-4=[10^{-5},10^{-4}]$ confirms that the two PNPs and Wegg’s potential share the GR slope, while the PW potential differs. However, there remains a noticeable offset between the PNP results and the exact GR precession: the $N_1=1$  potential deviates less  than $N_1=7$ for higher values of $L$. Both PNPs underperform relative to Wegg’s potential, which is nearly indistinguishable from GR over this range. Similarly, in the far‑field regime $L\gg 4$, the leading‑order precession correction scales as $\frac{6\pi}{L^2}$ (Equation \eqref{GRfarperc}), appearing as a linear relation in a log‑log plot. The linear graph intercept gives the correction amplitude ($6\pi$), meaning plots that display the correct behavior must coincide as $L$ grows. Panel [\ref{fig:Fig1}(III)] demonstrates that both PNPs correctly reproduce this behavior, while the PW potential does not. The PNP with $N_1=7$ performs slightly better than $N_1=1$ coinciding for lower values of $L$. Again Wegg’s potential closely matches the exact GR result (deviations remain under 1\% even at intermediate $L$, \cite{weggPSEUDONEWTONIANPOTENTIALSNEARLY2012a}) whereas the PNPs can deviate by up to 50\% outside the limiting cases. This is unsurprising, since the PNPs were constrained only at the two extremes and thus need not perform optimally elsewhere.

\begin{figure}[!]
    \centering
    \includegraphics[width=1\linewidth]{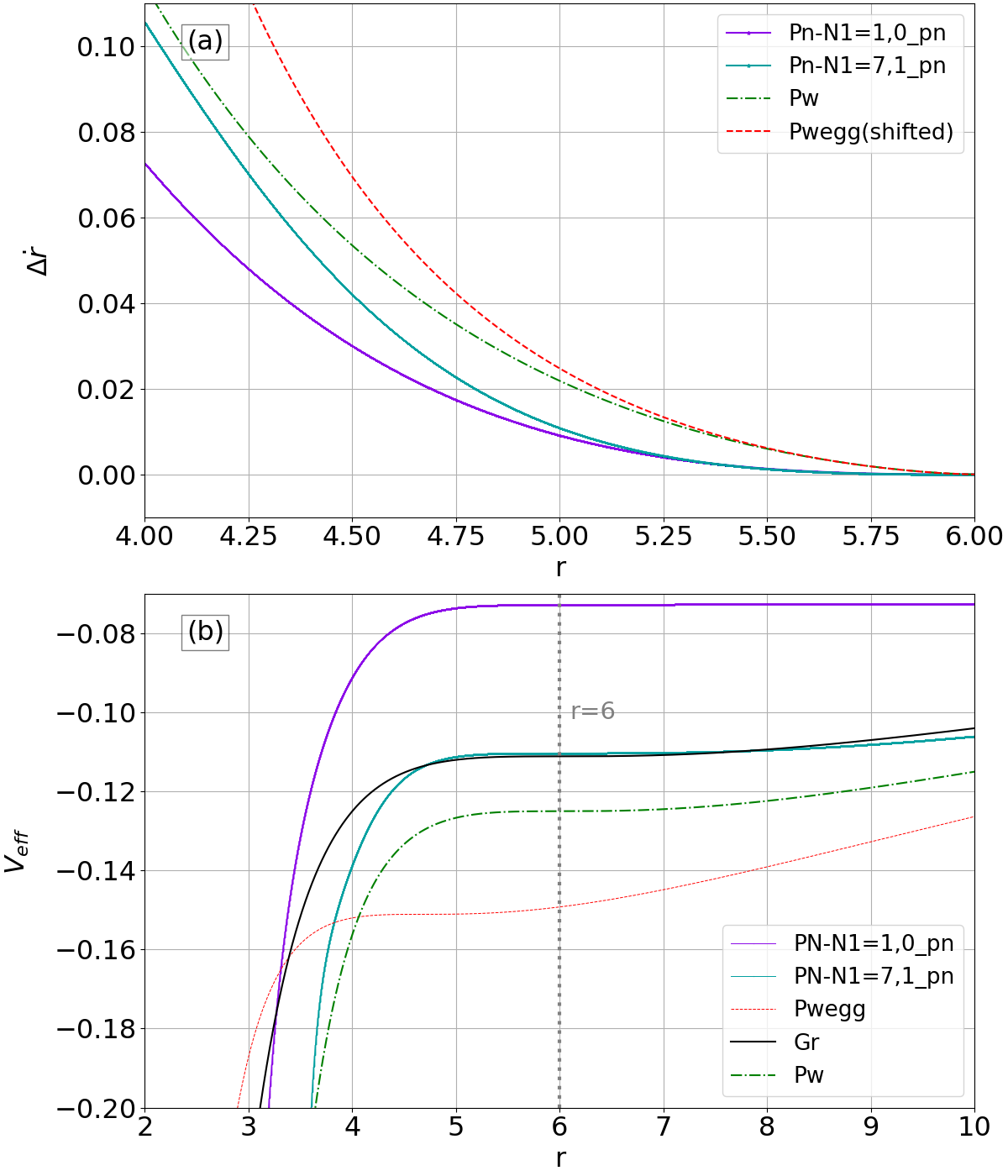}
    \caption{\footnotesize
    Comparison of two pseudo‑Newtonian potentials with $N_1=7$ (purple line) and $N_1=1$ (turquoise line), constructed using the constraints presented in sections \S\ref{sec:ISCO} and \S\ref{sec:prec}, alongside the PW potential (Equation  \eqref{PWp}, green dotted line), Wegg's potential (Equation \eqref{WeggPot}, red dashed line)  and the exact GR results (solid black line). (a) Deviation in radial velocity $\Delta\dot{r}=\dot{r}_{pnp}-\dot{r}_{gr}$ plotted against $r$, Between $r=4$ to the ISCO radius ($r=6$) on the GUI.  Wegg's potential result is shifted such that its ISCO aligns at $r=6$ for comparison. 
    (b) The Effective potential for $L_{ISCO}$ versus $r$,  highlighting the stationary inflection at $r=6$ for all curves besides Wegg's potential, and the monotonic decline toward $-\infty$ as $r\rightarrow 0$, indicative of the plunge. }
    \label{fig:Fig 2}
\end{figure}

Nevertheless, the proposed potentials offer additional advantages by reproducing further features of the exact GR geodesics, as shown in Figure [\ref{fig:Fig 2}].
Panel [\ref{fig:Fig 2}(a)] shows that, for radii close to the ISCO radius, the two PNPs exhibit smaller deviations in infall velocity relative to GR than either the PW or Wegg potentials. Moreover, this deviation  $\Delta\dot{r}$ approaches zero as $r\rightarrow 6$,  confirming that the new PNPs recover the correct plunge rate close to the ISCO. While the PW potential does reproduce the ISCO radius correctly, Wegg's potential does not: its ISCO lies at  $r_{ISCO}\approx4.7$, as shown in Panel [\ref{fig:Fig 2}(b)]. Consequently, Wegg’s radial‑velocity difference curve was shifted to  align its ISCO with $r=6$ for easier comparison. Both new PNPs not only recover the ISCO radius but also reproduce the initial infall velocity. Moreover, the  $N_1=7$ potential yields a more accurate specific orbital energy at the ISCO ($\approx 0.8\%$ deviation), as indicated by its effective potential at $r=6$  in Panel [\ref{fig:Fig 2}(b)].
As $r \rightarrow2$, however, the new PNPs deviate strongly from GR: the exact GR effective potential remains finite at the Schwarzschild radius, whereas the PNPs diverge by construction. For calculations requiring high accuracy near the event horizon, one could consider replacing the Schwarzschild radius in the denominator of the potential definition with a smaller value (In Equation \eqref{OurPot}).

TDE simulations provide a concrete use case, since orbital precession is essential for stream self-intersection and circularization. In both shallow penetration TDEs  ($r_- \gg r_{ISCO}$) and deep penetration ($r_- \approx r_{MBCO}$), the commonly used Paczyński–Wiita potential \citep{steinbergStreamDiskShocks2024} fails to  recreate the relevant precession rate, while our example potential reproduces the GR precession rate in these limits. Precession can be well matched by other simple potentials (e.g., Wegg’s), yet our example potential has improved dynamical behavior near $r_{ISCO}$, which becomes important when post-disruption accretion-disk formation and evolution are considered.

\section * {Conclusions:}
Two new PNPs were proposed as specific instances of the general form given in Equation \eqref{OurPot}. The suggested PNPs replicated seven features of Schwarzschild spacetime geodesics with great accuracy- specifically, the  precession per orbit in limiting cases, the radial velocity profile close to the ISCO radius on the GUI trajectory, and the MBCO and ISCO radii.  However, the proposed potentials performed rather poorly in replicating certain aspects such as the precession at intermediate values of angular momentum where Wegg’s potential performed better,  and the radial velocity as  $r\rightarrow 2$ on the GUI. 
\textbf{The proposed method and example potentials are not intended to replace full general relativistic magnetohydrodynamic simulations. Rather, for applications where pseudo-Newtonian treatments are used to reduce computational cost and complexity, this framework can systematically reduce the deviations from key general relativistic dynamical features while retaining the efficiency of PNPs, as shown above for the precession per orbit and the dynamics near the ISCO radius. This would be beneficial for TDE simulations. Moreover, the same linear-constraint construction allows straightforward system-specific “tailor-made” potentials when desired, with essentially no additional complications.}
Not every feature of Schwarzschild dynamics can be cast as a simple linear condition, and an arbitrary combination of constraints may yield an unsolvable system. Furthermore a set of equations could be solvable but result in an "over-fitted" potential, which could negate some of the simple conditions and damage the performance of the PNP. Nevertheless, this work demonstrates that several key geodesic properties can be matched by a straightforward PNP. Future work could explore this family of potentials further, either by changing the subtraction radii to a different constant  ($r_n=const$) or by considering the fully general form of the potential, and numerically determining its coefficients to identify even higher‑performance PNPs.

\section*{Appendix:}
\begin{table}[ht!]
    \centering
    \begin{tabular}{c|c|c}
        \textbf{Coefficient} & \textbf{$N_1=1$} & \textbf{$N_1=7$} \\
        \hline
        $\alpha_0$ & $\frac{226307703}{222470}\approx1017.25$ & $\frac{17832}{713}\approx25.01$ \\
        $\alpha_1$ & --- & $\frac{-17119}{713}\approx -24.01$ \\
        $\alpha_2$ & --- & $\frac{-33525}{713}\approx-47.02$ \\
        $\alpha_3$ & --- & $\frac{-172180}{713}\approx -241.49$ \\
        $\alpha_4$ & --- & $\frac{1453920}{713}\approx2039.16$ \\
        $\alpha_5$ & --- & $\frac{-14471560711}{999679}\approx -14476.21$ \\
        $\alpha_6$ & --- & $\frac{37215422990}{894783}\approx 41591.56$ \\
        $\beta_0$ & $\frac{-1073911624}{548153}\approx-1959.14$ &$\frac{-41804986773}{779528}\approx-53628.59$ \\
        $\beta_1$ & $\frac{1130048824}{545819}\approx 2070.37$ & --- \\
        $\beta_2$ & $\frac{-1263122286}{974125}\approx-1296.67$ & --- \\
        $\beta_3$ & $\frac{49145920}{102013}\approx481.76$ & --- \\
        $\beta_4$ & $\frac{-76518756}{777637}\approx-98.4$ & --- \\
        $\beta_5$ & $\frac{6915079}{810743}\approx 8.53$ & --- \\
        $\beta_6$ & $\frac{-199379605}{895302}\approx -222.69$ & --- \\
    \end{tabular}
    \caption{
    \textbf{The coefficients $\alpha$ and $\beta$ as defined in equation \ref{GenPot}, obtained by solving the system of equations defined by the conditions listed in the first paragraph of \S\ref{sec:exm}(Equations \ref{1stCond},\ref{2ndCond},\ref{4thCond},\ref{5thCond},\ref{6thCond},\ref{7thCond},\ref{8thCond},\ref{9thCond} ), meaning a total of $N_{tot}=8$ equations and coefficients, for the cases $N_1=7$ and $N_1=1$.}}
    \label{tab:coefficients_vertical}
\end{table}

\newpage

\bibliography{Bibliogrphy2025}
\bibliographystyle{aasjournal}

\end{document}